\documentclass{article}
\usepackage{times}
\usepackage{graphicx}
\usepackage{amsmath,amssymb}
\usepackage{a4wide}

\newcommand{\rmd}{{\rm d}}
\newcommand{\rme}{{\rm e}}
\newcommand{\rmi}{{\rm i}}

\begin{document}

\title{Interacting electrons in a magnetic field: mapping quantum mechanics to a classical ersatz-system}

%\classification{31.15.ac,31.15.Pf,73.21.La}
%\keywords      {interacting electrons in magnetic fields, quantum and classical Hall effects}

\author{Tobias Kramer\\
Institut f\"ur Theoretische Physik\\ 
Universit\"at Regensburg, 93040 Regensburg, Germany\\
Department of Physics, Harvard University, 02138 Cambridge, USA}

\maketitle

\begin{abstract}
Solving the quantum-mechanical many-body problem requires scalable computational approaches, which are rooted in a good understanding of the physics of correlated electronic systems. 
Interacting electrons in a magnetic field display a huge variety of eigenstates with different internal structures, which have been probed experimentally in the Hall effect.
The advent of high-performing graphics processing units has lead to a boost in computational speed in particular for classical systems.
In the absence of a quantum-computer, it is thus of importance to see how quantum-mechanical problems can be cast into a seemingly classical dynamics, which can be efficiently implemented.
At the same time, such mappings provide insights into the quantum-to-classical transition of many-body systems.
\end{abstract}

%\linenumbers

\section{Introduction}

The energy spectrum of interacting electrons in a magnetic field displays a rich structure, which results from the partial lifting of the degeneracies present in the non-interacting case.
The magnetic field case bears some similarities with the ordinary harmonic oscillator, but the coupling of the angular momentum to the direction of the magnetic field leads to a reshuffling of the oscillator energies into highly degenerate Landau levels \cite{Gottfried2004a}.
Once interactions between the particles are switched on, the oscillator states are no longer the eigenstates and in general numerical methods are required to obtain the energy-spectrum. 
Only for two-particles and specific Coulomb interaction values, exact analytic expressions are known \cite{Vercin1991,Taut1994,Kramer2010c}.
The effect of interactions in a quasi two-dimensional gas of electrons at low temperature has received an enormous interest due to the observation of the quantized Hall effect, which shows up in measurements as plateaus of the transverse resistivity across a Hall bar for certain ranges of magnetic fields.
The usual connotation of the integer and fractional Hall effects as being related to non-interacting and interaction electrons respectively, is somewhat arbitrary and questionable, since the electrons do not suddenly turn non-interacting in the integer Hall effect. 
Thus it is desirable to develop an understanding of both quantum Hall effects based on a common underlying Hamiltonian.
Somewhat surprisingly, the classical Hall effect discovered in 1879 \cite{Hall1879a} is presently not fully understood on a microscopic level. 
Recently, extensive numerical, first-principle calculations, became feasible using Graphics Computing Processors (GPUs), and have shown the importance of interactions and the proper incorporation of boundary conditions in the classical Hall effect in order to generate the observed Hall potential distribution \cite{Kramer2009c}. These computations require large particle numbers ($\sim 10^4$) to follow the self-consistent build-up of the Hall effect.
The computed distribution of the Hall potential is in excellent agreement with measurements performed by Knott et al.\ \cite{Knott1995a} and its possible implications for the quantum Hall effect are discussed in Ref.~\cite{Kramer2009b}.
In the converged solutions small regions next to the two contacts emerge, where the complete Hall potential drops over a tiny region in space. 
The continuum-solution based on conformal-mappings of the potentials contains a divergence of the electric field strength.
The existence of such singularities points to non-trivial physics near the contacts, since a linear-response theory (based on zero bias-voltage) cannot spontaneously generate such points for a non-zero bias voltage.
This motivates the search for possible similarities between the classical and quantum Hall effects and highlights the need to understand the importance of interactions in all incarnations of the Hall effects. 
\begin{figure}[t]
\begin{center}
\includegraphics[width=0.5\textwidth]{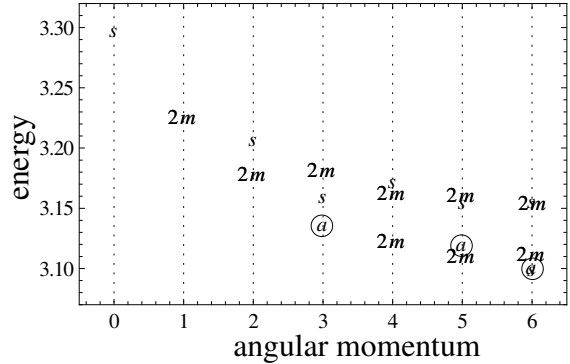}
\end{center}
\caption{Part of the spectrum of three spin-less electrons in a magnetic field sorted by angular momentum. The quasi two-dimensional electrons interact via Coulombic forces. The labels denote mixed (m), symmetric (s), and antisymmetric (a) symmetry of the spatial part of the wave function.
For fermionic (spin 1/2) electrons, the spatially symmetric states are not accessible. Energies are expressed in units of $\hbar\omega_l$.
\label{fig:spectrum}}
\end{figure}

Here, for simplicity, I will focus on model systems with fixed electron numbers and not discuss transport properties. Already the computation of the quantum-mechanical few-electron case is a difficult many-body problem. Direct diagonalization methods in harmonic oscillator basis-sets are limited to small particle numbers, and interest exists to develop scalable methods from few to several hundred electrons. For the fractional quantum Hall effect, Laughlin developed a scalable method to sample information about specific many-body states.
In the following, I review some known properties of Laughlin's wave-function and discuss how to use a classical map to compare the effect of the Pauli principle with the influence of the Coulomb interaction.

\section{Mapping of quantum mechanical systems to classical statistical mechanics}

One of my motivations is to discuss alternative methods to density functional theory (DFT), which employs a mapping of an interacting electron system to a non-interacting (still quantum-mechanical) Kohn-Sham system.
A DFT implementation has to resort to approximate functionals to achieve this mapping, since the exact functional is unknown. 
In addition DFT is restricted to calculating ground-state properties and does not easily yield insight into the internal structure of the wave function. Recently, density-matrix renormalization group (DMRG) methods have been employed to analyze electrons in magnetic fields \cite{Shibata2008}. DMRG yields some information about excited states and the internal structure.
A very different road is followed by methods which map the quantum-mechanical systems on classically looking equations. 
One example is coherent-state dynamics in combination with the time-dependent variational principle (TDVP), which allows one to derive systematically equations of motions for variational parameters of the wave function.
The basic idea is to store the state of a many-body wave function in terms of trajectories of the variational parameters, which for coherent states include the positions, momenta, and possibly the widths of the time-evolving wave function \cite{Kramer2010c,Grossmann2011a}.
The stationary properties of the system are recovered by taking a discrete number of realizations along a trajectory and use these as a basis set for a subsequent diagonalization \cite{Davis1979}. 
Laughlin proposed another approach, which maps a specific quantum state to a classical interacting system at finite temperature. In the following, I investigate how the incorporation of Coulomb interactions affects Laughlin's construction.
\begin{figure}[t]
\begin{center}
\includegraphics[width=0.29\columnwidth]{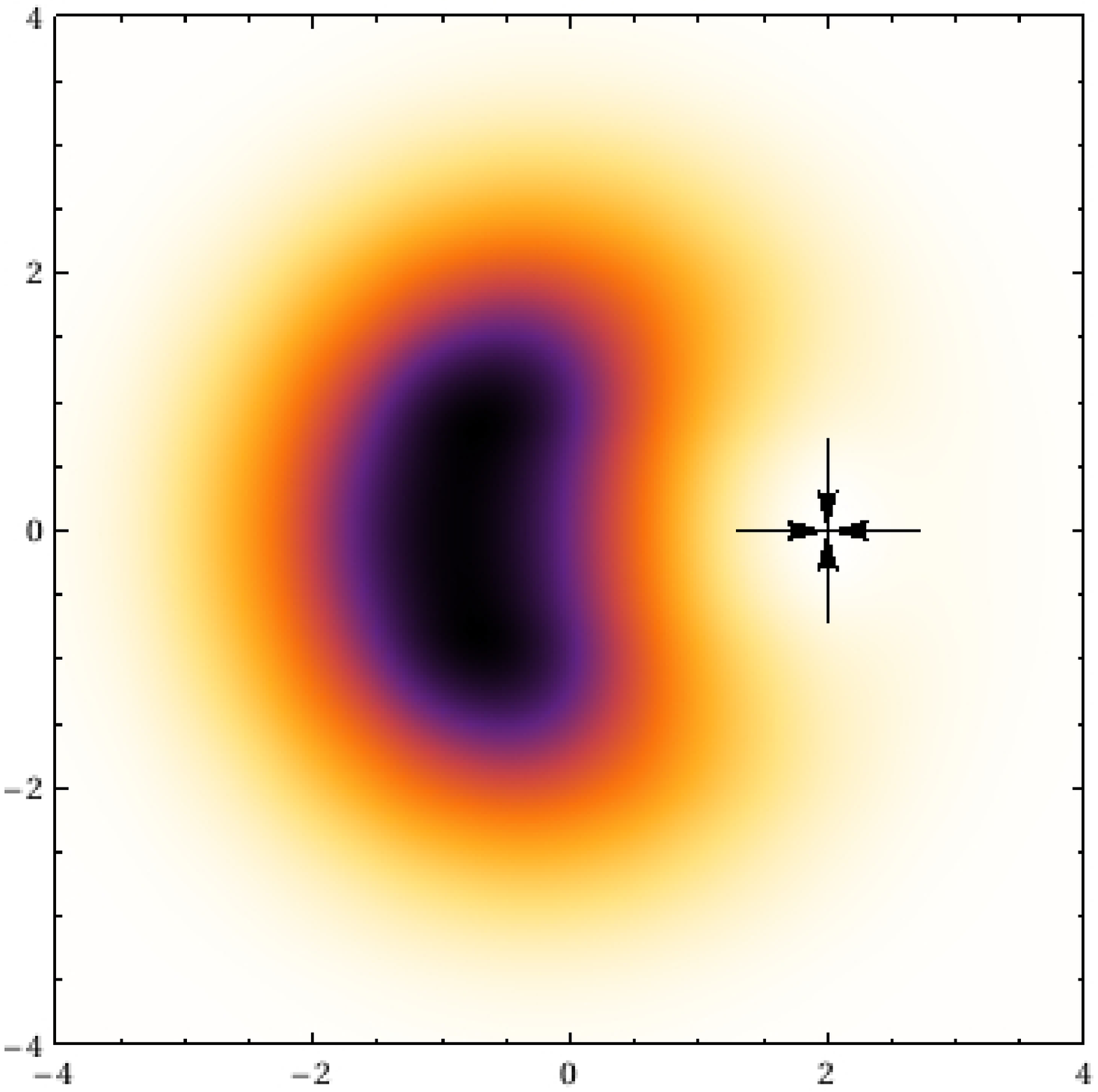}
\includegraphics[width=0.29\columnwidth]{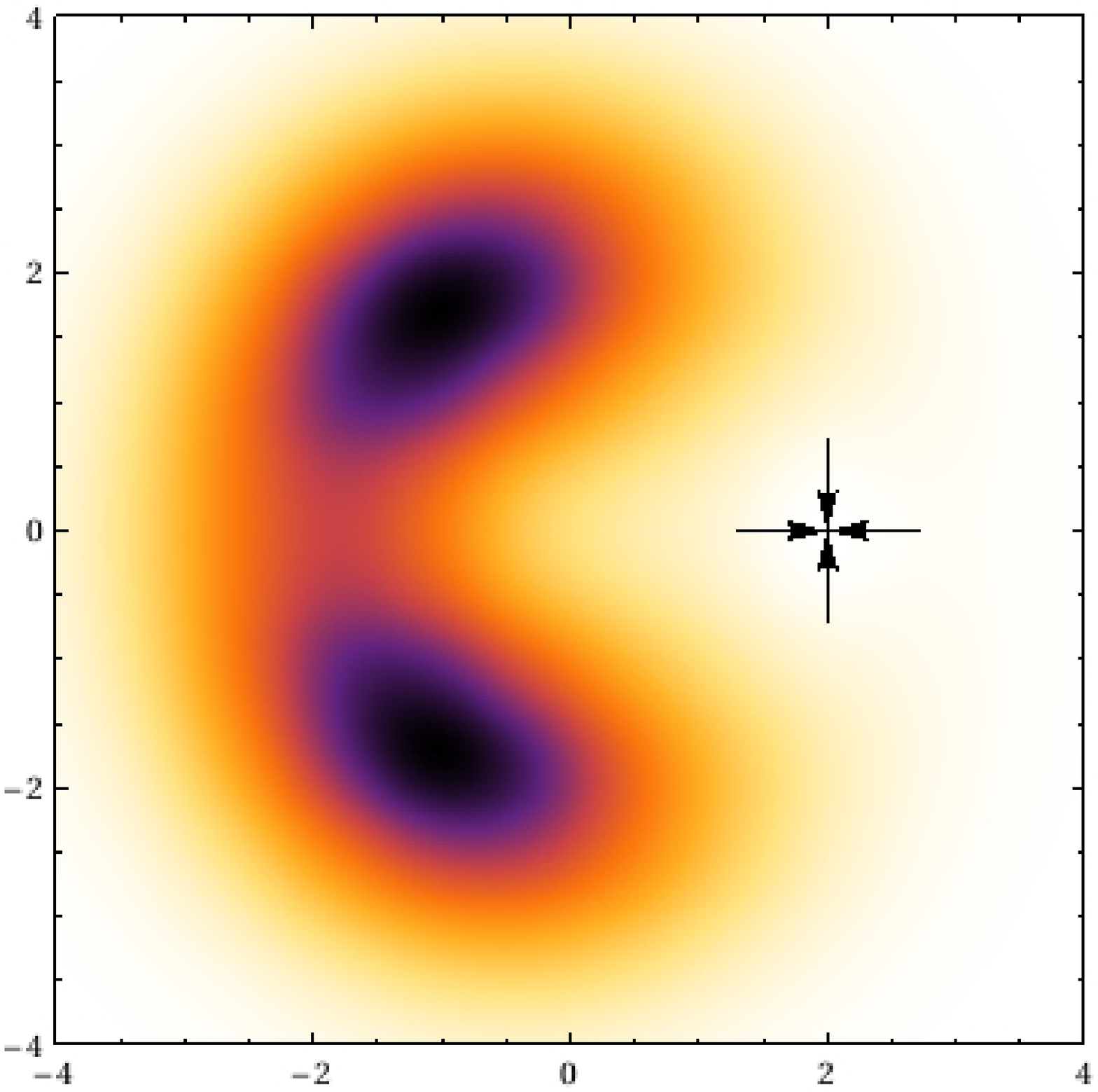}
\includegraphics[width=0.29\columnwidth]{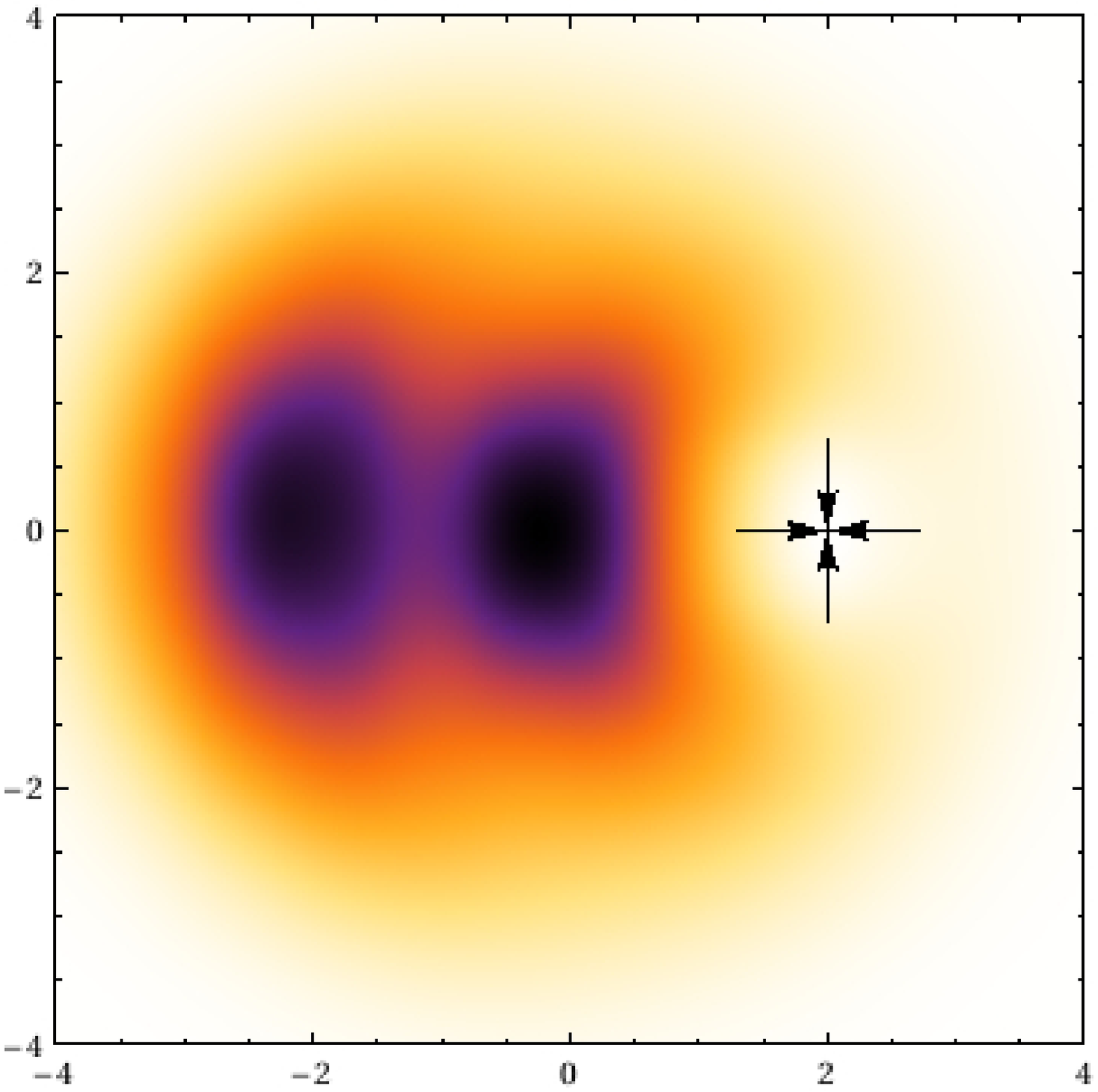}
\end{center}
\caption{Pair correlation function of three electrons for the antisymmetric states shown in Fig.~\ref{fig:spectrum}, from left to right panel $L=3,6,5$
\label{fig:corr}}
\end{figure}

\section{The three electron case}

As first example, I discuss the spectrum of three electrons put in a plane perpendicular to a uniform magnetic field. 
The spatial extension along the $z$-direction is neglected in the following and the Hamiltonian is given by
\begin{equation}
H_{\rm ee}=
\sum_{i=1}^3
\left[
\frac{\mathbf{p}_i^2}{2m}+\frac{1}{2}m\omega_l^2\mathbf{r}_i^2-\omega_l L_{z,i}
\right]
+\kappa\left( {|\mathbf{r}_1-\mathbf{r}_2|}^\alpha+{|\mathbf{r}_1-\mathbf{r}_3|}^\alpha+{|\mathbf{r}_2-\mathbf{r}_3|}^\alpha \right)
\end{equation}
with Larmor frequency $\omega_l=\frac{e B}{2m}$ and angular momentum 
$L_{z,i}=-\rmi\hbar(x_i \partial_{y,i}-y_i \partial_{x,i})$, $i=1,2,3$. 
The strength of the interaction is specified by $\kappa$. 
For Coulomb interaction I set $\alpha=-1$, $\kappa>0$, and for repulsive oscillator interactions $\alpha=2$, $\kappa<0$.
The Hamiltonian is separable into a center-of-mass part a relative part upon introducing Jacobi vectors
\begin{equation}
\begin{pmatrix}
\mathbf{s}_1\\
\mathbf{s}_2\\
\mathbf{s}_3
\end{pmatrix}
=
\begin{pmatrix}
\sqrt{1/2} & -\sqrt{1/2} & 0 \\
\sqrt{1/6} & \sqrt{1/6} & -2\sqrt{1/6} \\
\sqrt{1/3} & \sqrt{1/3} & \sqrt{1/3}
\end{pmatrix}
\begin{pmatrix}
\mathbf{r}_1\\
\mathbf{r}_2\\
\mathbf{r}_3
\end{pmatrix}.
\end{equation}
The center-of-mass part is contained in $\mathbf{s}_3$, and the relative part depends only on $\mathbf{s}_1,\mathbf{s}_2$. 
\begin{equation}
H_\text{rel}=
\sum_{i=1}^2
\left[
\frac{\mathbf{\tilde{p}}_i^2}{2m}+\frac{1}{2}m\omega_l^2\mathbf{s}_i^2-\omega_l L_{z,i}
\right],\quad
H_\text{cm}=
\frac{\mathbf{\tilde{p}}_3^2}{2m}+\frac{1}{2}m\omega_l^2\mathbf{s}_3^2-\omega_l L_{z,3}.
\end{equation}
The introduction of Jacobi coordinates greatly facilitates the analysis of the three-body Hamiltonian.
Especially the analysis of possible symmetries can be linked to the Jacobi vectors, as has been demonstrated by P.~Kramer and Moshinsky 
\cite{Kramer1966} and interaction matrix elements are simplified to integrals only involving two single-electron orbitals.
Simonovi\'{c} and Ciftja provide expressions for the exact Coulomb matrix elements of few-body systems \cite{Simonovic2006,Ciftja2009}.
Laughlin also analyses the three-electron problem with Jacobi vectors in chapter~7 of Ref.~\cite{Prange1987a}.
\begin{figure}[t]
\begin{center}
\includegraphics[width=0.4\textwidth]{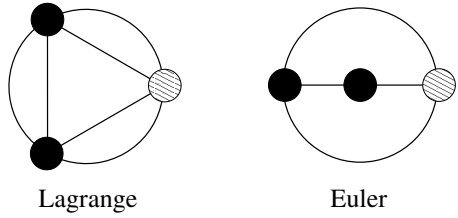}
\end{center}
\caption{Lowest configurations of three electrons underlying the pair-distribution function shown in Fig.~\ref{fig:corr}.\label{fig:conf}}
\end{figure}
The diagonalization of the relative part results in the spectrum shown in Fig.~\ref{fig:spectrum}. The angular-momentum of the relative motion is a good quantum number and for electrons in a magnetic field the distance between the electrons increases towards larger angular momenta. This in turn lowers the energy of the system.
The number of states with a specific symmetry for each angular momentum can be derived from \cite{Kramer1966}.
Iachello and coworkers also analyzed algebraic approaches to the three-body problem in relation to molecular vibrations \cite{Iachello1995,Bijker2000}.
To gain insight into the internal structure of the states, I show in Fig.~\ref{fig:corr} the pair-distribution function 
\begin{equation}
g_2(\mathbf{r}_1,\mathbf{r}_2)=\int\rmd \mathbf{r}_3 {|\psi(\mathbf{r}_1,\mathbf{r}_2,\mathbf{r}_3)|}^2
\end{equation}
for three antisymmetric states. The location of one electron is marked by the cross, while the shading indicates a high probability to find another electron. 
Two typical configurations underly these pictures, namely the triangular arrangement (Lagrange) and the colinear alignment (Euler) depicted in Fig.~\ref{fig:conf}.

\section{Laughlin's plasma map}

Laughlin proposed that states for $N$ electrons with coordinates $z_i=(x_i+\rmi y_i)/l_b$ of the form
\begin{equation}\label{eq:ls}
\psi_m(z_1,\ldots,z_N)=\prod^N_{i<j}{(z_j-z_k)}^m \exp\left\{-\frac{1}{4}\sum_{l=1}^N{|z_l|}^2 \right\}
\end{equation}
are of special robustness against the effect of interactions. The magnetic length is given by $l_b=\sqrt{\hbar/(2m\omega_l)}$.
The states are clearly antisymmetric under permutations of two particles and are states of angular momentum $\frac{N(N-1)m}{2}$.
These states are among the set of degenerate eigenstates of the non-interacting Hamiltonian ($\kappa=0$). Laughlin conjectured that the presence of interactions in general will (i) lift the degeneracies and (ii) will affect the family of states given by Eq.~(\ref{eq:ls}) less than all other states, thus leading effectively to a gap between the ground-state energy and the next excited state.
The Jastrow factor depends only on the relative coordinates, while the sum in the exponent can be viewed as a center-of-mass term, which is symmetric under exchange of two-particles.
Laughlin illuminates the properties of these states by noting that the probability density can be written as a partition function:
\begin{eqnarray}
\int\rmd z_1\cdots \rmd z_N\;{|\psi_m(z_1,\ldots,z_N)|}^2&=&\int\rmd z_1\cdots \rmd z_N\;\rme^{-\beta U_\text{cl}(z_1,\ldots,z_N)},\\
U_\text{cl}(z_1,\ldots,z_N)&=&-2m^2\sum_{j<k}\ln|z_j-z_k|+\frac{m}{2} \sum_j {|z_j|}^2\\
\beta&=&2m.
\end{eqnarray}
\begin{figure}[t]
\begin{center}
\includegraphics[width=0.48\columnwidth]{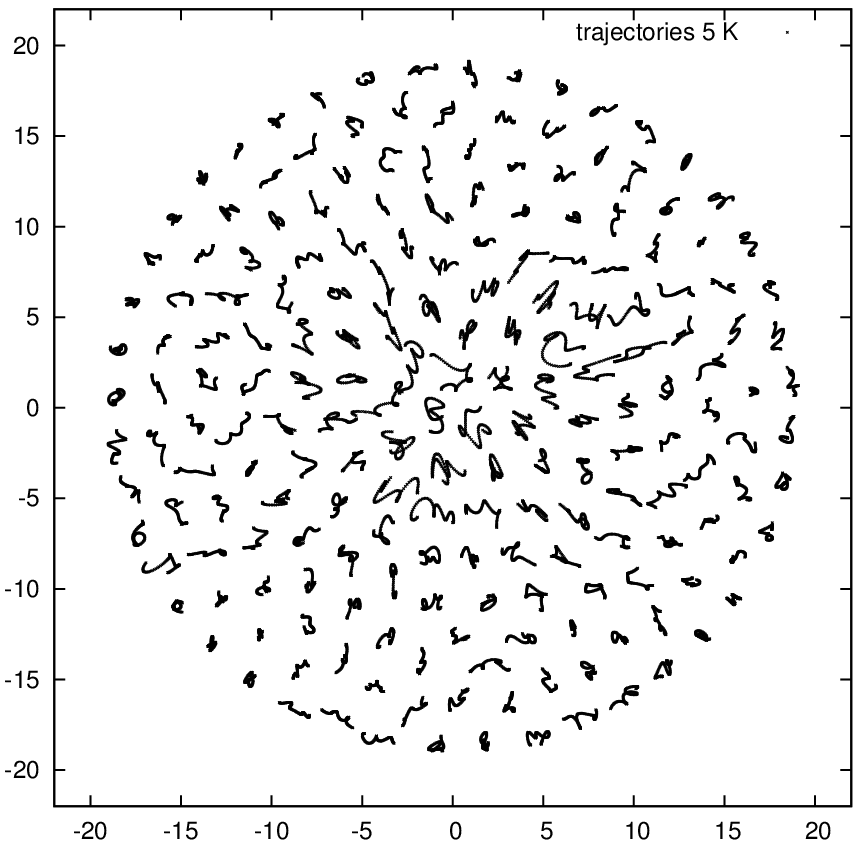} ~~~
\includegraphics[width=0.48\columnwidth]{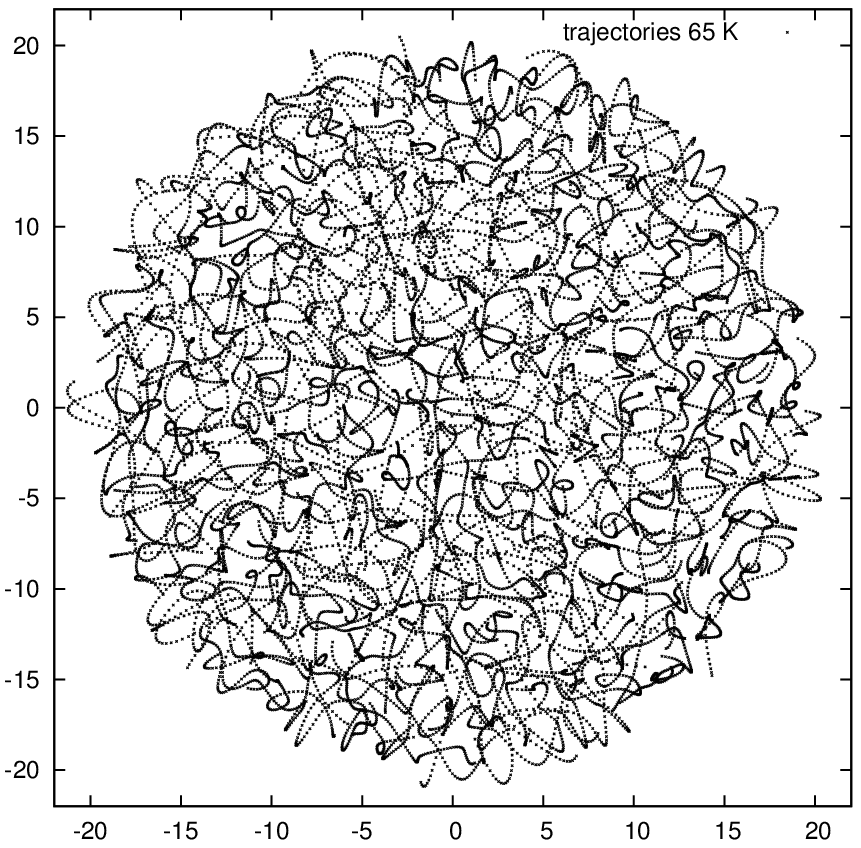}
\end{center}
\caption{Trajectories of the classical ersatz-dynamics for $m=1$. Left panel: low temperature ($T=5$~K), right panel: temperature $T=65$~K corresponding to $\beta=2m$. 
By evaluating the histogram of the particle positions and distances, the density and pair correlation function of the equivalent quantum-mechanical state are evaluated.
\label{fig:wurm}}
\end{figure}
It is important to realize that this mapping to a classical system transforms the Fermi-Dirac statistics of the quantum-mechanical state at zero temperature to a classical Boltzmann system at temperature $1/(k_B T)=2m$. 
The Pauli principle, which hinders two electrons to occupy the same state, is encoded in the logarithmic classical interaction of two point charges. 
A smooth density is achieved by translating the ``fuzziness'' of the quantum-mechanical location of the electrons to the thermal motion in the classical ersatz-system.
For instance, randomly distributed charges at initial time $t=0$ will start to move and collide and convert part of their electrostatic energy into kinetic energy. 
After a short while, an equilibration between kinetic and electrostatic energies is reached, which allows one to determine the temperature of the system by looking at the histogram of the kinetic energy and to compare it to the Boltzmann distribution $\exp(-\beta E_\text{kin})$.
Only for $\beta=2m$ do the classical density distribution $n(\mathbf{r})$ and pair correlation function $g_2(\mathbf{r1},\mathbf{r2})$ coincide with their quantum mechanical counterparts. 
Fig.~\ref{fig:wurm} shows a visualization of the mapped state for $m=1$, where electrons are at time $t=0$ randomly distributed and then a classical propagation using Newton's equations of motion is performed and the trajectories are plotted. The left panel shows the classical system at a too low temperature, while in the right panel the particles have the correct temperature to reproduce the statistics of $|\psi_1(\mathbf{z}_1,\ldots,\mathbf{z}_{196})|^2$. 
\begin{figure}[t]
\begin{center}
\includegraphics[width=0.4\textwidth]{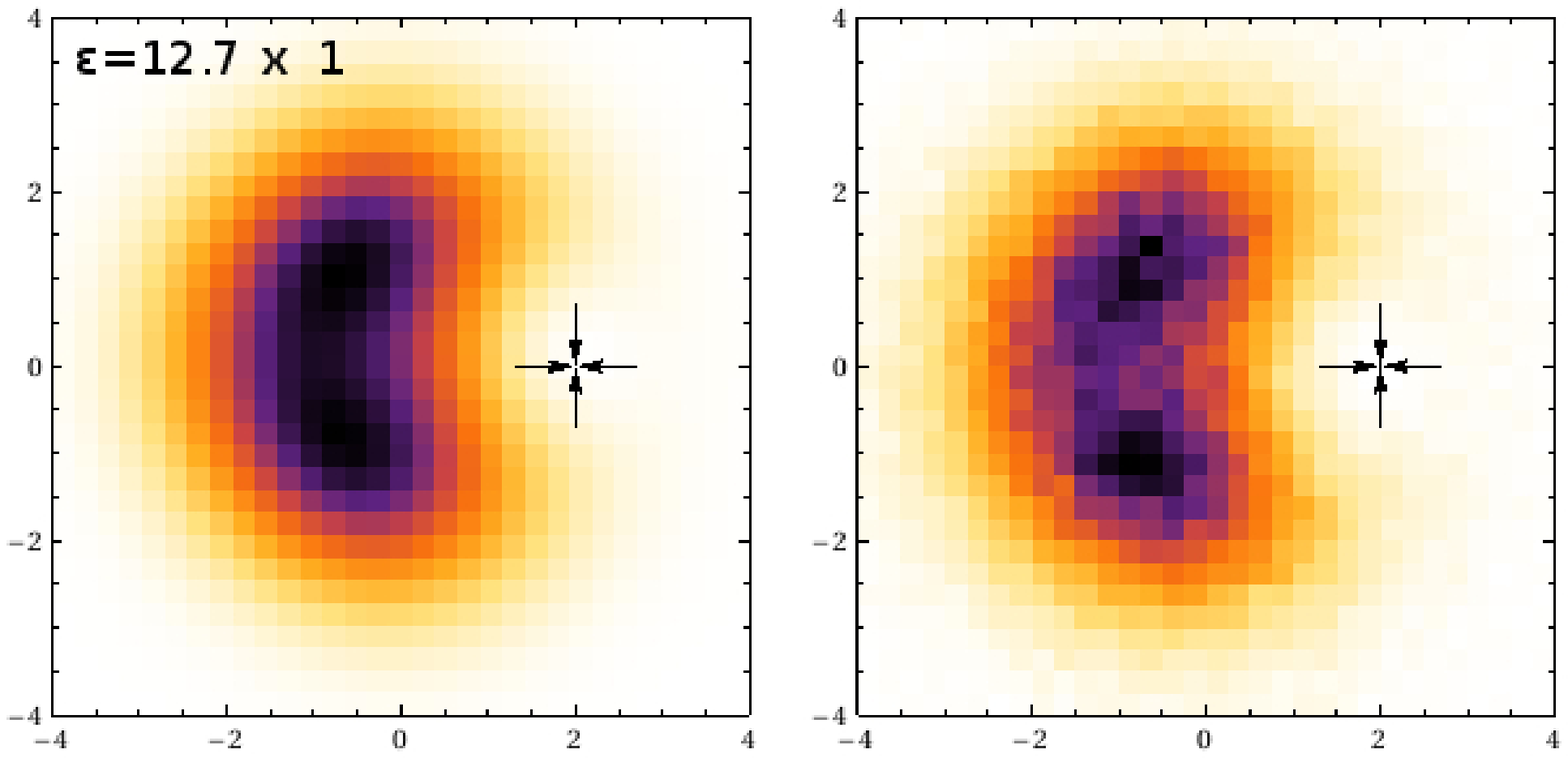}~~~~\includegraphics[width=0.4\textwidth]{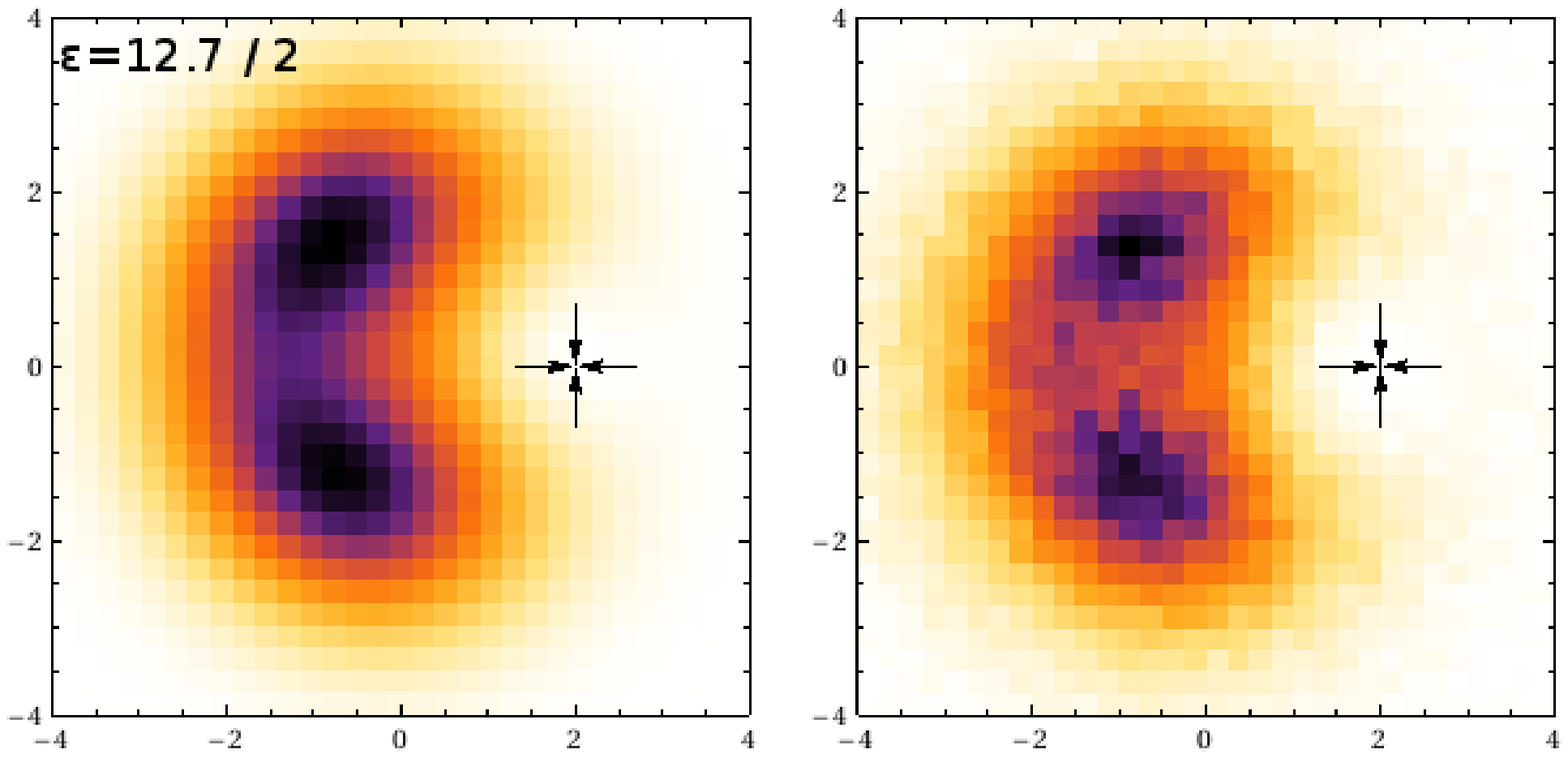}
\end{center}
\caption{Pair correlation function for three Coulombic-interacting electrons in a magnetic field.
Left two panels: weak Coulomb interaction, left-most panel: exact diagonalization, second panel histogram of the pair-distribution of the classical ersatz-dynamics using Eq.~(\ref{eq:pc}). 
Right two panels: increased Coulomb interaction, exact diagonalization vs. ersatz-dynamics (right-most panel).
\label{fig:paircorr}}
\end{figure}
Several methods are available to determine the pair-correlation function: Laughlin used the hypernetted-chain method \cite{Prange1987a} and the Metropolis algorithm is discussed in Ref.~\cite{Ciftja2003}.
Here, I will use a direct integration of the classical equations of motions and obtain the density and pair correlation function by sampling the electron positions periodically.
In the classical ersatz-dynamics, the effect of the Coulomb interaction is included by amending the classical potential containing the logarithmic ``Pauli-potential'' with a Coulombic term:
\begin{equation}\label{eq:pc}
U_\text{cl,Coulomb}(\mathbf{R})=-2m^2\sum_{j<k}\bigg[\underbrace{\ln|z_j-z_k|}_{\text{Pauli}}+\underbrace{\kappa {|z_j-z_k|}^{-1}}_{\text{Coulomb}}\bigg]+\frac{m}{2} \sum_j {|z_j|}^2
\end{equation}
In order to fill out phase-space for the three electron case, an ensemble average over different initial positions of the electrons is performed.
In Fig.~\ref{fig:paircorr} I compare the pair correlation function of the mapped system with the result from an exact diagonalization of the Hamiltonian.
The main effect of the Coulomb interaction is a stretching of the pair-correlation function, seen as an increase in the relative distances between the electrons.

\section{Larger systems}

\begin{figure}[t]
\begin{center}
\includegraphics[width=0.3\textwidth]{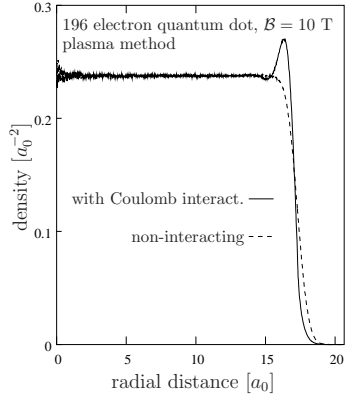}
\end{center}
\caption{Radial density profile of an 196 electron quantum dot, including a disk-shaped positive background charge. The dashed line shows the non-interacting case, which represents the lowest Landau level ($m=1$). The solid line denotes the result including the Coulomb term in Eq.~(\ref{eq:pc}). DFT results for similar systems are shown in \cite{Pi1998}.
\label{fig:196}}
\end{figure}
The three-electrons in a magnetic field were an important test-case for Laughlin's development of a theory of the fractional quantum Hall effect. It is desirable to investigate how the classical ersatz-dynamics works for larger particle numbers. This has been investigated by Monte-Carlo samplings \cite{Ciftja2003}. In addition to the Coulomb forces, in this case also a uniform, disk-shaped, positive background is taken into account. Here, I present results using again a direct integration of Newtons equations for 196 electrons, which are in good agreement with the results shown in Ref.~\cite{Ciftja2003}. The integration is performed very efficiently using GPU computers. Similar systems have been investigated using current density-functional theory \cite{Pi1998,Lipparini2008}.

\section{Summary}

I have reviewed some aspects of electrons in a magnetic field. The classical ersatz-dynamics provides an intriguing representation of the Pauli principle in the form of an interacting electron gas at finite temperature. The addition of Coulomb forces yields good results for Laughlin's states already for three electrons. 
For more general electronic systems, the classical-ersatz dynamics has seen some renewed attention, see for instance the overview in Ref.~\cite{Dharma-wardana2011}.
If the ersatz-dynamics provides a viable alternative to other many-body methods such as DFT, DMRG, and TDVP, requires further investigation and a careful comparison of the results for different systems.
One difficulty of Laughlin's map is its reliance on a specific state, 
rather than the ability to obtain a whole set of eigenstates.
From a computational point of view, the classical dynamics can be excellently implemented on the emergent super-computing hardware consisting of graphics processing units.

\section*{Acknowledgments}

I thank Roelof Bijker and Alejandro Frank for the invitation to present this work at the Cocoyoc meeting in honor of Prof.~F.~Iachello. Financial support from the Emmy-Noether program of the Deutsche Forschungsgemeinschaft (grant KR 2889/2) is gratefully acknowledged. 
I am indebted to Peter Kramer for helpful discussions and comments.

\providecommand{\url}[1]{#1}

\end{document}